\documentclass[11pt,a4paper]{article}
\usepackage{amsmath}
\usepackage[psamsfonts]{amssymb}
\usepackage{amsmath}
\usepackage{epsfig}
\usepackage{graphicx}

\title{Super-acceleration in non-minimal derivative coupling model}
\author{H. Mohseni Sadjadi\footnote{mohsenisad@ut.ac.ir},
\\{\small Department of physics, University of Tehran,}
\\{\small P.O.B. 14395-547, Tehran, Iran}.}

\begin{document}
\maketitle

\begin{abstract}
A scalar field model with non-minimal derivative coupling to
gravity is considered. It is shown that although in the absence of
matter and potential the phantom divide line crossing is
forbidden, but for the power law potential and in the presence of
matter this crossing is, in principle, possible.
\end{abstract}

\section{Introduction}

Recent astrophysical data indicate that the expansion of our
universe is accelerating \cite{ac}. If the universe is assumed to
be filled with perfect fluids, and we adopt the Einstein theory of
gravity, nearly  $\%70$ of our universe must be filled by a smooth
component with a negative equation of state parameter $w<-1/3$,
dubbed as dark energy, giving rise to a negative pressure. A
natural simple candidate for dark energy, inspired by the scalar
field model of inflation, is a dynamical scalar field with a
suitable potential, known as the quintessence \cite{quint}.
Although this model can describe the accelerated expansion, but is
unable to explain the phantom divide line crossing which based on
some observations may be occurred in recent era \cite{cross}. A
model consisting of only one scalar field can not describe this
crossing, hence another model, comprising a quintessence field and
a phantom scalar field (a scalar field with a wrong sign kinetic
term \cite{phant}), known as quintom model was
introduced\cite{quintom}.

By generalizing the models including non-minimal coupling of
scalar field to gravity, and by assuming that the coupling terms
may also be functions of derivatives of the scalar field, the
inflationary cosmological dynamics was discussed in \cite{AM}.
Late time acceleration of the universe as well as the inflation
era may be studied in the framework of such non-minimal coupling
models.

When the scalar field is kinetically coupled to the Einstein
tensor, the field equation is of the second order and no new
degrees of freedom are introduced. The cosmological applications
of this model were investigated in \cite{sus1}, and exact
solutions were obtained in the absence of matter and potential. In
\cite{sus2}, the results of \cite{sus1} were extended to include
the phantom scalar field and also constant potentials. However,
finding an exact analytical solution to Friedmann equations in
non-minimal derivative coupling model, even for simple kinds of
potential, and in the presence of matter is very complicated, and
it seems that is not feasible analytically \cite{sus2}. More
generalized models (with additional couplings) and some special
cosmological solutions corresponding to power law expansion,  both
in the late and early time stages of the FLRW universe were also
studied in \cite{grand}.  Extension  of non-minimal derivative
coupling  model to dynamical gravity and aspects of Hawking
radiation in this model can be found in \cite{Haw}.

Recently, to reconcile the slowly rolling scalar field (inflaton)
with the Higgs boson, the same model in which the unitary bounds
are preserved has been used \cite{germ}.

In this paper by considering a spatially flat FLRW universe and
inspired by the model proposed in \cite{germ}, we consider a
slowly rolling scalar field in non-minimal derivative coupling
model. We show that besides the acceleration expansion of the
universe, this model can describe the super acceleration without
using new additional scalar fields. The scheme of the paper is as
following:

In the first part of the second section we introduce the model and
show that in the absence of the potential and matter, the scalar
field can not super accelerate the universe. In the second part by
considering a power law potential, and also by taking the matter
contribution into the account, we prove that, in principle,
solutions describing the phantom divide line crossing may exist in
this model. In the third section we conclude the paper.

We use units $\hbar=c=G=1$ throughout the paper.

\section{Phantom divide line crossing in non-minimal derivative coupling model}
We consider a scalar field model with the action
\begin{equation}\label{1}
S=\int \left({R\over 16\pi}-{1\over 2}g^{\mu
\nu}\partial_\mu\varphi \partial_\nu\varphi+{k\over 2}G^{\mu \nu
}\partial_{\mu}\partial_{\nu}\varphi-V(\varphi)\right)\sqrt{-g}d^4x+S_m,
\end{equation}
where $G^{\mu \nu}(=R^{\mu \nu}-{1\over 2}g^{\mu \nu}R)$ is the
Einstein tensor, and $S_m$ is the action of the matter sector.  We
study the model in a spatially flat FLRW space time equipped with
the metric:
\begin{equation}\label{2}
ds^2=-dt^2+a^2(t)(dx^2+dy^2+dz^2).
\end{equation}
The Friedmann equations are
\begin{equation}\label{3}
H^2={8\pi\over 3}\left({\dot{\varphi}^2\over
2}+V(\varphi)+{9k\over 2}H^2\dot{\varphi}^2\right)+{8\pi \over
3}\rho_m,
\end{equation}
which may be written as
\begin{equation}\label{4}
H^2={8\pi \over 3}\left({{\dot{\varphi}^2\over
2}+V(\varphi)+\rho_m\over 1-12\pi k \dot{\varphi}^2}\right),
\end{equation}
and
\begin{equation}\label{5}
\dot{H}=-4\pi\left({\dot{\varphi}^2+V(\varphi)+\rho_m\over 1-12\pi
k \dot{\varphi}^2}+ {\dot{\varphi}^2-V(\varphi)+P_m\over 1-14\pi k
\dot{\varphi}^2}\right)+8\pi k {H\dot{\varphi}\ddot{\varphi}\over
1-4\pi k \dot{\varphi}^2}.
\end{equation}
The scalar field equation and the continuity equation for the
matter component with energy density $\rho_m$, are given by
\begin{eqnarray}\label{6}
&&(1+3kH^2)\ddot{\varphi}+\left(3H(1+3kH^2)+6kH\dot{H}\right)\dot{\varphi}+V'(\varphi)=0,
\nonumber \\
&& \dot{\rho_m}+3H\gamma_m\rho_m=0.
\end{eqnarray}
In terms of the equation of state parameter of the barotropic
matter, $w_m$, $\gamma_m$ is defined by $\gamma_m=w_m+1$. We have
assumed that there is no interaction between the scalar field and
the matter component. Friedmann equations can be rewritten as
\begin{eqnarray}\label{8}
H^2&=&{8\pi \over 3}(\rho_d+\rho_m) \nonumber \\
\dot{H}&=&-4\pi (P_d+\rho_d+P_m+\rho_m),
\end{eqnarray}
where the effective dark component pressure and energy density are
defined trough
\begin{equation}\label{9}
P_d={{{\dot{\varphi}^2\over
2}-V(\varphi)-2kH\dot{\varphi}\ddot{\varphi}+4\pi
k\dot{\varphi}^2P_m\over 1-4\pi k\dot{\varphi}^2}},
\end{equation}
and
\begin{equation}\label{10}
\rho_d={{{\dot{\varphi}^2\over 2}+V(\varphi)+12\pi
k\dot{\varphi}^2\rho_m\over 1-12\pi k\dot{\varphi}^2}}
\end{equation}
respectively. The continuity equation for this effective dark
component is
\begin{equation}\label{11}
\dot{\rho_d}+3H(P_d+\rho_d)=0.
\end{equation}
In contrast to the ordinary scalar field model, the matter
contributes in $P_d$ as well as in $\rho_d$. For $k=0$, where
\begin{equation}
\dot{H}=-4\pi (\dot{\varphi}^2+\gamma_m\rho_m)<0,
\end{equation}
we are left with the usual scalar field model which can not cross
the phantom divide line characterized by $w=-1$ ($w={P\over
\rho}=-1-{2\over 3}{\dot{H}\over H^2}$ is the equation of state
parameter of the universe).  But as we will prove, the
modification of the kinetic term via non minimal coupling, may
enable this crossing (provided that some conditions be satisfied).
Let us first investigate the ability of this model to cross $w=-1$
for the most simple case where $V=0$ and the contribution of
matter, $\rho_m$, is negligible (i.e. a universe dominated by free
$\varphi$ component \cite{sus1}). By defining
$X=(1+3kH^2)\dot{\varphi}$, one can verify that the field equation
(\ref{6}) leads to the autonomous equation
\begin{equation}\label{12}
{dX\over d\ln a}=-3X,
\end{equation}
with fixed points located at $1+3kH^2=0$ and $\dot{\varphi}=0$. By
solving (\ref{12}) we get
\begin{equation}\label{13}
\dot{\varphi}={Ca^{-3}\over 1+3kH^2},
\end{equation}
where $C$ is a constant. (\ref{13}) and (\ref{4}) imply
\begin{equation}\label{14}
a^{-6}={3\over 4\pi C^2}{(1+3kH^2)^2\over 1+9kH^2}.
\end{equation}
By taking time derivative of both sides of (\ref{14}), we arrive
at
\begin{equation}\label{15}
\dot{H}={-3H^2(1+3kH^2)(1+9kH^2)\over 1+9kH^2+54k^2H^4}.
\end{equation}
This is the same as what was obtained in \cite{sus1} (see eq.(22)
of  \cite{sus1}) using a different mathematical procedure. This
autonomous equation has no real pole and its fixed points are
located at $H=0$, $kH^2=-{1\over 3}$, and $kH^2=-{1\over 9}$. It
is clear from (\ref{14}) that $kH^2>-{1\over 9}$, therefore
$\dot{H}<0$ and $\lim_{t\to \infty}{\dot H}=0$, in agrement with
the results of \cite{sus1}, \cite{sus2}. In this way, $\dot{H}=0$
cannot be crossed and only may be achieved asymptotically. So
there is no acceleration to super-acceleration phase transition
for the universe in this model.

In the following we investigate if the presence of the matter and
the potential can change this situation and super-accelerate the
universe. Solving Friedman equations, for a non constant potential
even in the absence of matter is a complicated task which based on
our knowledge has not yet be done. To get an insight about what
happens in the presence of the potential and the matter and to see
if the quintessence to phantom phase transition is allowed in this
model, let us take the power law potential
\begin{equation}\label{16}
V(\varphi)=v\varphi^n,\,\,\,\, v\in \Re,
\end{equation}
and try a {\it possible} slow roll solution specified by
\begin{equation}\label{17}
\ddot{\varphi}\ll 3H\dot{\varphi},
\end{equation}
and \cite{sad}
\begin{equation}\label{18}
H=h_0+h_1t^\alpha,\,\,\,\alpha(\geq 2)\in {N}.
\end{equation}
If it is found that (\ref{18}) satisfies the Friedman equations
with an even integer $\alpha$ and $h_1>0$, then the transition
from quintessence phase to phantom phase at $t=0$ is allowed. Note
that (\ref{18}) can be considered as the Taylor series of $H$
around $t=0$, thence $h_1$ is proportional to the first non zero
derivative of $H$ at $t=0$ and $\alpha$ is the degree of this
derivative. Among (\ref{4}), (\ref{5}), and (\ref{6}), two
equations are independent. To see if (\ref{18}) is a solution, we
focus on equations (\ref{5}), and (\ref{6}).

For slowly varying field (\ref{17}) the field equation reduces to
\begin{equation}\label{19}
\left(3H(1+3kH^2)+6kH\dot{H}\right)\dot{\varphi}+nv\varphi^{n-1}=0.
\end{equation}
By substituting (\ref{18}) in (\ref{19}), and in terms of
dimensionless quantities $\tilde{h_1}:={h_1\over h_0^{\alpha+1}}$
, $\tilde{v}:={v\over h_0^2}$, $\tilde{k}:=kh_0^2$, and
$\tau:=h_0t$ we obtain
\begin{equation}\label{20}
\left(3(1+3\tilde{k})+\alpha\tilde{k}\tilde{h_1}\tau^{\alpha-1}+\mathcal{O}(t^{\alpha})\right){d\varphi\over
d\tau}+\tilde{v}\varphi^{n-1}=0.
\end{equation}
The solution of this equation is
\begin{equation}\label{21}
\varphi(\tau)=\left({n(n-2)\tilde{v}\tau\Phi\left(-{6\alpha
\tilde{h_1}\tilde{k}\over 3(1+3\tilde{k})
}\tau^{\alpha-1},1,{1\over \alpha-1}\right)\over
3(1+3\tilde{k})(\alpha-1)}+D\right)^{-{1\over n-2}}.
\end{equation}
$\Phi$ is the Lerchphi function defined by
\begin{equation}\label{22}
\Phi(x,a,b)=\sum_{n=0}^{\infty}{x^n\over (n+b)^a},
\end{equation}
and $D$ is determined by the value of the scalar field at $t=0$,
$D=\varphi^{2-n}(0)$. In the continue, for the sake of
mathematical simplicity we assume $\tilde{k}\gg1$. As was
mentioned in \cite{germ} this is the condition for canonical
normalization of the field $\varphi$. To see if this solution
satisfies the Friedmann equation we proceed as follows:

By expanding  $\mathcal{K}:=\dot{\varphi}^2$, and
$\tilde{V}:={V\over h_0^2}$ at $\tau=0$ and after some computation
we acquire
\begin{eqnarray}\label{23}
&&\tilde{\mathcal{K}}:={\mathcal{K}\over h_0^2}=
\tilde{\mathcal{K}}_0+\left(-{8\over
3}\tilde{h_1}\tilde{\mathcal{K}}_0-{18(n-1)\over
n}{\tilde{k}}{\tilde{\mathcal{K}}_0 ^2\over
\tilde{V_0}}\right)\tau +\mathcal{O}(\tau^2)\nonumber \\
&&\tilde{V}=\tilde{V}_0-9\tilde{k}\tilde{\mathcal{K}_0}\tau+\mathcal{O}(\tau^2),
\end{eqnarray}
where the subscript "0" denotes the value of the parameter at
$t=0$ and
\begin{equation}\label{24}
\tilde{\mathcal{K}}_0={n^2\varphi(0)^{2(n-1)}\tilde{v}^2\over 81
\tilde{k}^4}.
\end{equation}
The matter density also can be expanded at $\tau=0$ to give
\begin{equation}\label{25}
\tilde{\rho}_m:={\tilde{\rho}_{m}\over
h_0^2}=\tilde{\rho}_{m0}-3\gamma_m \tilde{\rho}_{m0}\tau+
\mathcal{O}(\tau^2).
\end{equation}
By substituting (\ref{23}), (\ref{25}), and (\ref{18}) in the
Friedmann equation, (\ref{5}), we obtain :
\begin{eqnarray}\label{26}
&&\alpha\tilde{h_1}\tau^{\alpha-1}+\mathcal{O}(\tau^\alpha)=-4\pi{\mathcal{P}}\nonumber \\
&-&4\pi\left({\mathcal{A}\left(-{8\over
3}\tilde{h}_1\tilde{\mathcal{\mathcal{K}}}_0-{18(n-1)\over
n}{\tilde{k}}{\tilde{\mathcal{K}}_0^2\over
\tilde{V}_0}\right)-3\mathcal{B}
\gamma_m\tilde{\rho}_{m0}-9\mathcal{C}\tilde{k}\tilde{\mathcal{K}}_0}\right)\tau\nonumber
\\
&&+\mathcal{O}(\tau^2),
\end{eqnarray}
where $\mathcal{P}$, $\mathcal{A}$, $\mathcal{B}$ and
$\mathcal{C}$ are defined through
\begin{equation}\label{27}
\mathcal{P}={\tilde{\mathcal{K}}_0(1-8\pi k\mathcal{K}_0+8\pi k
V_0)+\gamma_m\tilde{\rho}_{m0}-4\pi (1+3w)k\tilde{\mathcal{K}}_0
\rho_{m0}\over (12\pi k \mathcal{K}_0-1)(4\pi k \mathcal{K}_0-1)},
\end{equation}
\begin{eqnarray}\label{28}
\mathcal{A}&=&{1+4\pi k(w+3)\rho_{m0}+8\pi k V_0(1-48\pi^2
k^2\mathcal{K}_0^2)-16\pi k \mathcal{K}_0(1-5\pi k
\mathcal{K}_0)\over
(12\pi k \mathcal{K}_0-1)^2(4\pi k \mathcal{K}_0-1)^2}\nonumber \\
&-&{96\pi^2 k^2
\rho_{m0}\mathcal{K}_0(\gamma_m+2\pi(1+3w)k\mathcal{K}_0)\over
(12\pi k \mathcal{K}_0-1)^2(4\pi k \mathcal{K}_0-1)^2},
\end{eqnarray}
\begin{equation}\label{29}
\mathcal{B}={2w-16\pi wk\mathcal{K}_0(3-22\pi k \mathcal{K}_0+48
\pi^2 k^2 \mathcal{K}_0^2)\over (12\pi k \mathcal{K}_0-1)^2(4\pi k
\mathcal{K}_0-1)^2},
\end{equation}
and
\begin{equation}\label{30}
\mathcal{C}={8\pi k \mathcal{K}_0(1+48 \pi^2 k^2
\mathcal{K}_0^2-16\pi k \mathcal{K}_0)\over (12\pi k
\mathcal{K}_0-1)^2(4\pi k \mathcal{K}_0-1)^2}
\end{equation}
respectively.

Validity of (\ref{26}) requires: i) $\alpha=2$, ii)
$\mathcal{P}=0$, and iii)
\begin{equation}\label{31}
\tilde{h_1}=-2\pi\left({\mathcal{A}\left(-{8\over
3}\tilde{h}_1\tilde{\mathcal{\mathcal{K}}}_0-{18(n-1)\over
n}{\tilde{k}}{\tilde{\mathcal{K}}_0^2\over
\tilde{V}_0}\right)-3\mathcal{B}
\gamma_m\tilde{\rho}_{m0}-9\mathcal{C}\tilde{k}\tilde{\mathcal{K}}_0}\right).
\end{equation}

On the other hand by inserting  (\ref{21}) in (\ref{17}), we find
that the conditions for validity of the approximation (\ref{17})
in the neighborhood of the transition time ($t=0$) are
\begin{equation}\label{32}
\tilde{h}_1={h_1\over h_0^3}\ll 1,\,\,\,{d^2\tilde{V}\over
d\varphi^2}\ll \tilde{k},
\end{equation}
which results in $n(n-1)\tilde{v}\varphi(0)^{n-2}\ll \tilde{k}$
implying $\mathcal{K}_0\ll V_0$. One can also make use of
(\ref{4}) to obtain $k \mathcal{K}_0\ll 1$. With the help of these
relations, $\mathcal{P}=0$ reduces to
\begin{equation}\label{33}
\gamma_{m}\rho_{m0}\simeq -8\pi k\mathcal{K}_0 V_0.
\end{equation}
Therefore we must take $k<0$. Besides, from (\ref{4}) or (\ref{8})
we find out that the main contribution in $H^2$ is coming from the
potential: $h_0^2\simeq {8\pi \over 3}V_0$.

By considering $\tilde{k}\tilde{\mathcal{K}}_0=k \mathcal{K}_0$
and $kV_0\tilde{\mathcal{K}}_0\ll 1$, from (\ref{31}) we acquire
\begin{equation}\label{34}
\tilde{h}_1\simeq 2\pi\left(72\pi ({3n-2\over n}) k^2
\mathcal{K}_0^2+6\gamma_m w_m \tilde{\rho}_{m0}\right).
\end{equation}
Now let us investigate the conditions required for
$\tilde{h}_1>0$. If $w\neq 0$, then  $\tilde{h}_1\simeq
12\pi\gamma_m w \tilde{\rho}_{m0}>0$, and the transition occurs.
If $w_m=0$, as is expected when $\rho_m$ is taken as the cold dark
matter, $\tilde{h}_1>0$ requires $n>{2\over 3}$ or $n<0$. These
results show that the transition from quintessence to phantom
phase is in principle allowed in this model. At the end let us
note that by virtue of (\ref{33}), the main contribution of the
energy density is coming from the scalar field potential, which
implies $\rho_{m0}\ll \rho_{d0}$. To obtain a same order of
magnitude for $\rho_m$ and $\rho_d$,  and to alleviate the
coincidence problem one should consider an interaction between
these components.

\section{Conclusion}

We investigated whether phantom divide line (defined by $w=-1$,
where $w$ is the equation of state parameter of the universe)
crossing can be realized in a scalar field model whose the kinetic
term is non-minimally coupled to gravity (see (\ref{1})). We began
our study with a simple model without the matter and the scalar
field potential, and showed that super-acceleration is not
possible in this situation and $w=-1$ can be only achieved
asymptotically (see the discussion after (\ref{15})). In the next
step we considered the contribution of matter (such as cold dark
matter or ordinary matter with positive pressure) and took the
power law potential for the scalar field (see (\ref{16})). As in
this case, obtaining an analytical exact solution to Friedmann
equations is not possible, we tried an approximate slowly rolling
solution (see(\ref{17}) and (\ref{18})) and showed that the
transition from quintessence to phantom phase may be in principle
occurred in this model.

\end{document}